# Positron Emission Tomography (PET) image enhancement using a gradient vector orientation based nonlinear diffusion filter (GVOF) for accurate quantitation of radioactivity concentration

Dr Mahbubunnabi Tamal*

Department of Biomedical Engineering

Imam Abdulrahman Bin Faisal University, Dammam, Saudi Arabia

***Corresponding Author:**
mtamal@yahoo.com, mtamal@iau.edu.sa

## ABSTRACT

To accurately quantify in vivo radiotracer uptake using Positron Emission Tomography (PET) is a challenging task due to low signal-to-noise ratio (SNR) and poor spatial resolution of PET camera along with the finite image sampling constraint. Furthermore, inter lesion variations of the SNR and contrast along with the variations in size of the lesion make the quantitation even more difficult. One of the ways to improve the quantitation is via post reconstruction filtering with Gaussian Filter (GF). Typically GF is implemented with varying kernel widths accounting for the intrinsic resolutions of different scanners at the expense of reduction in spatial resolution. Edge preserving Bilateral Filter (BF) and Nonlinear Diffusion Filter (NDF) are the alternatives to GF that can improve the SNR without degrading the image resolution. However, the performance of these edge preserving methods are only optimum for high count and low noise cases. A novel parameter free gradient vector orientation based nonlinear diffusion filter (GVOF) is proposed in this paper that is insensitive to statistical fluctuations (e. g., SNR, contrast, size etc.). The performance of the method is compared with GF, BF and NDF using SNR, contrast-to-noise ratio (CNR), resolution, percentage bias and reproducibility for different count levels and lesion to background contrasts (2:1 and 4:1) with NEMA torso phantom. The GVOF method provides the highest average SNR 40.73 followed by NDF, BF and GF with average SNR 38.98, 23.52 and 21.33 respectively. There is almost an 100% increase in the average CNR for the GVOF method compared to the original unfiltered image (114.81 vs. 3.65). The CNR for NDF, BF and GF methods are 100.40, 23.11 and 9.88 respectively. The GVOF method also improves resolution by 10% compared to the original unfiltered image (average resolution 5.78 vs. 6.35 mm Full Width at Half Maximum). The average resolutions of NDF, BF and GF methods are 6.95, 6.47 and 8.86 mm. The percentage bias in estimating the maximum activity representing $SUV_{max}$ (Maximum Standardized Uptake Value – an image based biomarker) for the spheres with diameter > 2cm where the partial volume effects (PVE) is negligible is the lowest for the GVOF method with average bias of -1.40% (range -7.06 to 6.21%) followed by NDF 10.25% (-9.02 to 28.68%), GF 18.40% (6.78 to 33.82%), BF 61.23% (13.22 to 111.52%) and no filter 75.62% (19.29 to 124.69%) across different sizes, contrasts and acquisition durations. The GVOF method also improves the maximum intensity reproducibility by 106%, 5%, 157% and 60% for contrast 2:1 and 247%, 53%, 326% and 104% for contrast 4:1 compared to no filter, GF, BF and NDF methods respectively. Since the proposed GVOF method provides the most accurate quantitation of radioactivity concentration measured by the PET camera along with the improved resolution, SNR and CNR and robust across different sizes, contrast levels and acquisition durations, it can be used for both accurate diagnosis and response assessment for lesion size > 2cm without PVE correction. On the other hand, because of its capability to provide accurate quantitative measurements irrespective of the SNR, it can also be effective in reduction of radioactivity dose.

**KEYWORDS**

**INTRODUCTION**

Radiotracer uptake at molecular level measured by the Positron Emission Tomography (PET) scanner is useful for detection, diagnosis and staging of diseases [1, 2]. With the continuous development of new tracers, changes in molecular level tracer uptake recently have been used as a biomarker to monitor early response to therapy [3-5]. Standardized uptake value (SUV) – which is a normalized relative measure of the radioactive concentration to remove the variation of injected radio activity and patient size [5, 6], is widely used as a semiquantitative parameter for clinical evaluations [3, 7]. SUVs derived from PET images are not robust and reproducible due to poor signal-to-noise ratio (SNR) and low spatial resolution of PET camera [2]. The uncertainties in longitudinal SUV measurements are even greater when SNR, size and contrast of lesion are subject to change [8].

To achieve satisfactory signal-to-noise (SNR) ratio of PET images, filtering can either be performed during the image reconstruction process by incorporating a prior term in the statistical reconstruction algorithm [9] or after the completion of reconstruction known as post reconstruction filtering [10, 11]. Conventionally, Gaussian Filter (GF) with varying levels of full width half maximum (FWHM) accounting for the intrinsic resolutions of different scanners are employed [2, 12, 13]. Though GF provides linear and predictable SNR facilitating identification and delineation of lesions, it reduces the spatial resolution of images and thus accurate quantitation of radiotracer concentration [2, 14, 15].

The other alternatives to GF are edge preserving Bilateral Filter (BF) [16] and Nonlinear Diffusion Filter (NDF) [17]. Though through the incorporation of both spatial and intensity information [18] the BF improves resolution, its performance are not optimal for low SNR PET. The NDF method uses gradient magnitude information to calculate the diffusion coefficient. Due to high noise and limited resolution, the determination of gradient of PET images are not straightforward specially for low contrast and low count cases. NDF also known to increase noise instead of eliminating it if the image is corrupted by speckle noise [19].

Wavelet based filtering [20] is capable of smoothing based on the magnitude of noise. However, the method is still limited by capacity of processing edges discontinuities [21]. A combination of wavelet and curvelet based denoising method has recently been proposed to apply on PET images which complement the limitations of each other [22]. A hybrid spatial-frequency domain filtering is proposed recently that utilize the advantages of both spatial and transform domain filtering [23]. In this method, the non-local mean algorithm is considered as spatial domain filter for denoising high contrast region and the multi-scale curvelet approach for denoising the low contrast features. Though all these methods are capable of preserving edges, the quantitative accuracy are sometimes compromised. A unsupervised deep learning method has recently been proposed to smooth PET images and shows superior performance compared to other conventional methods [24]. However, the performance of the method is highly dependent on the training data and because of this, the performance may vary with different statistical settings.

Ideally, an optimal image filtering method should improve the SNR while preserving the original structure, geometric and quantitative information [22]. Since in response assessment, noise (i.e., the radiotracer uptake), contrasts and size can change, the method should also be robust enough to be implemented for different noise levels and contrasts without the need of adjusting the parameters for each scenario.

In this study, a new gradient vector orientation based nonlinear diffusion filter (GVOF) is proposed to meet the above criteria. The coefficient is calculated utilizing both the gradient magnitude and the gradient vector orientation information within a fixed neighborhood. The results of the new filter are compared with GF, BF and NDF at different noise levels and contrasts using a standard NEMA torso phantom.

**THEORY**

OVERVIEW OF DIFFUSION FILTER

The linear homogeneous and isotropic diffusion process is described by

$$\frac{\partial I}{\partial t} = \nabla \cdot [c \nabla I] \qquad (1)$$

The diffusion coefficient c has a constant value in the given domain, $\nabla \cdot$ and $\nabla$ are the divergence and gradient operators respectively. Perona and Malik [17] introduced non-linear isotropic diffusion process into image processing, where the diffusion coefficient $c(x, y, t)$ is dependent on the image structure

$$\frac{\partial I(x, y, t)}{\partial t} = \nabla \cdot [c(x, y, t) \nabla I(x, y, t)]$$

$$I(t = 0) = I_0 \qquad (2)$$

where $I(x, y, t)$ represents an image, $(x, y)$ represents a position of pixel in the image, t is a parameter indicating a scale level in a scale space of the image $I(x, y, t)$, $I_0$ represents an initial image with $t = 0$, $\nabla.$ and $\nabla$ are divergence and gradient operators respectively and $c(x, y, t)$ is the non-linear diffusion function which determines the behaviour of the diffusion process at a pixel located at position $(x, y)$. The purpose of the function $c(x, y, t)$ is to stop the diffusion in the image at the regions of object boundaries. Perona and Malik [17] also suggested the following choice for function $c(x, y, t)$

$$c(x, y, t) = \exp\left[-\left(\frac{\|\nabla I_\sigma(x,y,t)\|}{\kappa}\right)^2\right] \qquad (3)$$

where $\|\nabla I_\sigma(x, y, t)\|$ is the gradient magnitude of the original image $I(x, y, t)$ at current iteration t smoothed with a Gaussian kernel of width, $\sigma$. This smoothing operation can avoid oscillations of $\nabla I_\sigma(x, y, t)$ caused by signals in a noisy image and the smoothing is applied only once on the original noisy image. $\kappa$ controls the steepness of smoothing. Several other diffusivity functions have been suggested by different groups based on either the regularized version of the image to calculate the gradient [25] or the image gradient histogram [26]. All these non-linear filters have been designed and used to eliminate noise and preserve (enhance) the object edges. However, they fail to eliminate noise when the gradient magnitude is not evaluable and can even enhance speckle noise instead of eliminating it [19].

PROPOSED DIFFUSION FILTER

A gradient vector orientation based nonlinear diffusion filter (GVOF) is proposed in this work to overcome the limitations of the existing methods arising due to low contrast and high noise such as in the case of low resolution and low count PET. The central idea of the GVOF method is that the orientation of the gradient vectors are systematic within a small neighborhood at the boundary but random for noise as shown in Figure 1 and described in US Patent Application US 15/964,428 [27].

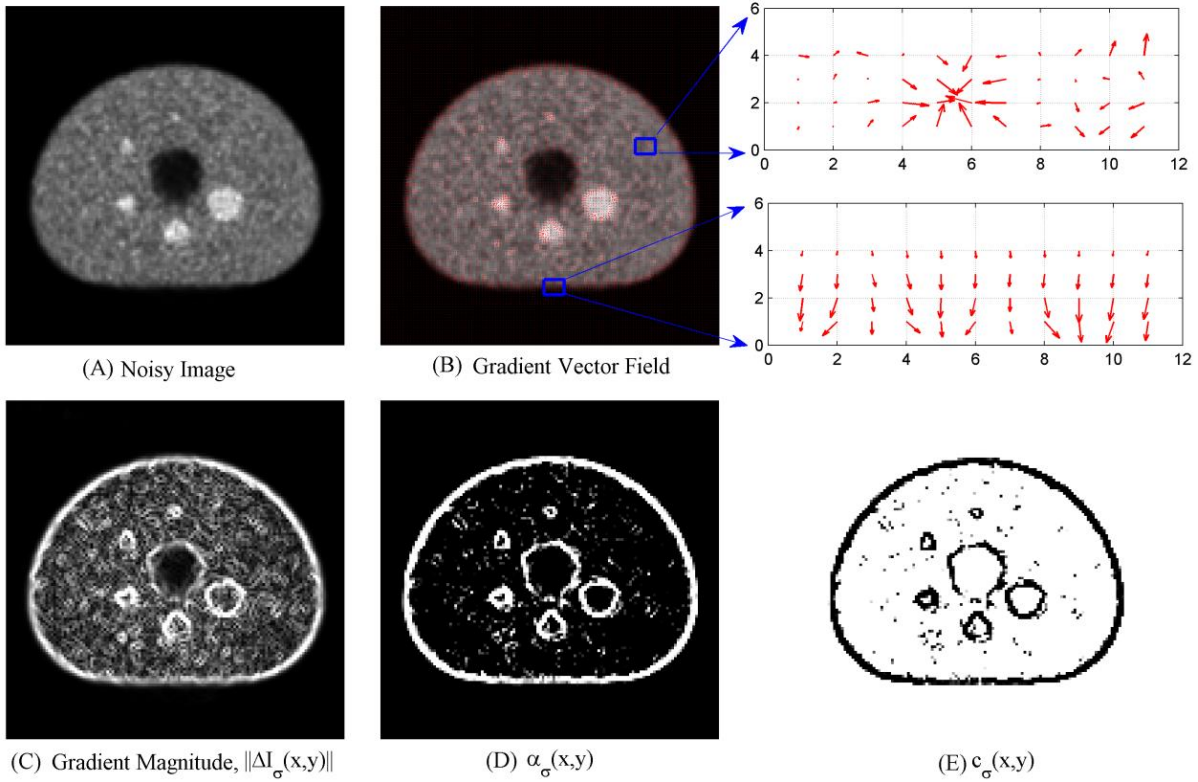

Figure 1: Illustration of the proposed method. At the edge of an object, the directions of the neighbouring gradient vectors are oriented in the similar direction as shown in (A). the direction of the gradient vectors in the middle of the object due to noise is randomly oriented as shown in (B). Gradient magnitude, cosine angles of the gradient vectors $\alpha_\sigma(x, y)$ and the proposed diffusion coefficient are shown in (C), (D) and (E) respectively.

The calculation of diffusion coefficient is thus dependent both on the magnitude and the direction of the gradient vectors within a specified window as given below

$$c(x, y, t) = \exp\left[-\left(\frac{\|\nabla I_\sigma(x,y,t)\| \times \alpha_\sigma(x,y,t)}{\kappa}\right)^2\right] \quad (4)$$

where the new term $\alpha_\sigma(x, y, t)$ is the sum of the cosine angles of the gradient vectors $\nabla I_\sigma$ of the smoothed image within the window and calculated as

$$\alpha_\sigma(x, y, t) = \sum_{i=1}^{p} \sum_{j=1}^{q} \cos(\theta_{(i,j,t)}) = \sum_{i=1}^{p} \sum_{j=1}^{q} \frac{\nabla I_\sigma(x,y,t) \cdot \nabla I_\sigma(i,j,t)}{\|\nabla I_\sigma(x,y,t)\| \|\nabla I_\sigma(i,j,t)\|} \quad (5)$$

$$\alpha_\sigma(x, y) = \frac{\alpha_\sigma(x,y) - \min(\alpha_\sigma)}{\max(\alpha_\sigma) - \min(\alpha_\sigma)} \quad (6)$$

where p and q defining the size of the window and (i, j) is the index of the pixel within the window. An angle between the gradient vector $\nabla I_\sigma(i, j, t)$ and the gradient vector $\nabla I_\sigma(x, y, t)$ can be represented as $\theta_{(i,j,t)}$. Thus, as shown in equation (5), the term $\alpha_\sigma(x, y, t)$ equals to a value of a sum of cosines $\cos(\theta_{(i,j,t)})$ of angles $\theta_{(i,j,t)}$ (that correspond to each gradient vector $\nabla I_\sigma(i, j, t)$ with respect to the gradient vector $\nabla I_\sigma(x, y, t)$). $\alpha_\sigma(x, y, t)$ is calculated for each 2D slice. Equation (6) is used to restrict the minimum and maximum values of $\alpha_\sigma(x, y, t)$ between to 0 and 1, with $\alpha_\sigma(x, y, t) = 0$ being indicative of randomness in the gradient orientations. In such case, the image is heavily smoothed. The method iteratively smooth the image. It can be run until there is no significant change between two subsequent images or can be run for fixed number of iterations. The pseudo code of the algorithm is described below:

**F = GVOF (I)**

**Input:**
    **I:** PET image;

1. **$I_\sigma$**: Filter image with Gaussian kernel of 4mm FWHM;
2. Select the first slice of **$I_F$** = one slice of **$I_\sigma$**
3. Iteration:
    For t=1:1:fixed iteration (or until there is no significant change in subsequent iteration)
        **c(x, y, t)**: Find proposed diffusion co-efficient using Equation 3
        **$I_F$**: Apply Non-linear diffusion filter to the slice and update
    End
4. Repeat for all slices.
5. **F:** stack slices to create a filtered volume

**Output:**
    **F:** Filtered image.

MATERIALS AND METHODS

PHANTOM DATA ACQUISITION

The torso NEMA phantom containing six spheres with diameter of 10, 13, 17, 22, 28 and 37 mm corresponding to the volume of 0.52, 1.15, 2.57, 5.58, 11.49 and 26.52 ml was filled with $^{18}$F solutions to yield two different contrasts between the hot spheres and the colder uniform background (2:1 and 4:1). The activity ratio between spheres and background are shown in Table 1.

Table 1: Activity Concentration

|  | 2:1 (KBq/ml) | 4:1 (KBq/ml) |
| --- | --- | --- |
| **Sphere** | 1668 | 2775 |
| **Background** | 838 | 697 |
| **Measured Ratio** | 1.99:1 | 3.98:1 |

The phantom data were acquired in a 3D mode on the TrueV PET-CT scanner (Siemens, USA) for 120 minutes which provides 109 image planes or slices covering a 21.6 cm axial FOV (field of view). Images were reconstructed into a 256×256×109 matrix with voxel dimensions of 2.67×2.67×2.00 mm using OSEM reconstruction algorithm with 4 iterations and 21 subsets for five different scan durations (900, 1200, 2000 and 4000 seconds [15, 20, 33.3 and 66.6 minutes respectively]) to represent different levels of noise. The starting time of each static frame were shifted to reconstruct five different non-overlapping and overlapping realizations for all durations. Full details of the data acquisition procedure are available in [28]. All the reconstructed images were smoothed with the proposed filter after applying decay correction and the performance of the new filter were then compared with the decay corrected images smoothed with a 4-mm FWHM (full width at half maximum) Gaussian filter, bilateral filter (4-mm and 20 for spatial and intensity domain respectively) and nonlinear diffusion filter ($\kappa = 0.5$ and 10 iterations). For the proposed GVOF method 60 iterations were chosen with κ=0.1. The parameters for all these filters were empirically chosen to provide the best results across different PET images.

SNR of the image was measured by placing a fixed size ROI of 26.52 ml in the background away from the hot sphere to minimize the spill in and partial volume effects and defined according to [29]

$$\text{SNR} = 20\log_{10}\left(\frac{\mu_{ROI}}{\sigma_{ROI}}\right) \quad (7)$$

where $\mu_{ROI}$ and $\sigma_{ROI}$ are the mean and standard deviation (SD) within the ROI.

Contrast-to-noise ratio (CNR) is used to evaluate the contrast as function of noise defined as the ratio of the difference of signal intensities of two regions of interest to the background noise [30, 31]

$$\text{CNR} = \frac{\mu_{SP} - \mu_{BG}}{\sigma_{BG}} \quad (8)$$

where $\mu_{SP}$ and $\mu_{BG}$ are the mean activity within the sphere and background. For estimating $\mu_{SP}$, a ROI was first estimated for each sphere using the calculated boundaries based on the known diameter and position of each sphere and eroded by 3×3 structural element to remove the partial volume effect. $\mu_{BG}$ is estimated by placing a fixed size ROI of 26.52 ml in the background away from the hot sphere similar to the SNR measurement. Since CNR indicates how visible or detectable the hot sphere is in contrast to the background within which the hot sphere is located, it is also known as detectability [32, 33].

Resolution as full width at half maximum (FWHM) was derived from the standard deviation of a 1D Gaussian function that was fitted to the absolute gradient of the edge profile of each image.

For assessing the accuracy of measurement of activity concentration after smoothing with different filters, percentage bias is calculated using the mean of the maximum activity of five realizations ($AC_{max}^{mean}$) within each sphere and true measured activity concentration (TAC) that was used to fill each sphere and is given by

$$\% \text{ Bias} = 100 \times \left(\frac{AC_{max}^{mean} - TAC}{TAC}\right) \quad (8)$$

$AC_{max}$ represents $SUV_{max}$ without correction for weight and injected dose.

The reproducibility of $AC_{max}$ was measured as the percent difference between the two measurements [34] and is calculated by considering the highest and lowest activity amongst five realizations within each sphere ($AC_{max-highest}$ and $AC_{max-lowest}$ respectively). The percentage difference equation in this case is given by

$$\% \text{ difference} = 100 \times \left[\frac{AC_{max-highest} - AC_{max-lowest}}{\frac{1}{2} \times (AC_{max-highest} + AC_{max-lowest})}\right] \quad (9)$$

**RESULTS**

Once slice of the original reconstructed images along with the images after application of different fileting methods are shown in Figure 2 for visual analysis. From Figure 2, it is clearly evident that the GF method improves the image quality by removing noise. Nonetheless the images appear blurred. The BF method provides images with higher resolution compared to the GF for 2000 and 4000 seconds acquisition duration. However, for 900 and 1200 seconds acquisition durations, quality of the images filtered by the BF method is inferior compared to the GF method. The NDF and GVOF method provide visually better quality images compared to all the other methods. The quality of the images are similar for 2000 and 4000 seconds for these two methods. For 900 and 1200 seconds acquisition durations, images smoothed with the proposed GVOF method appear superior to the NDF.

Profiles through the red lines A and B (bottom left image in Figure 2) for 900 and 4000 seconds for all methods are shown in Figure 3. Profile A passes through the 22 mm and 37 mm spheres, whereas profile B only passes through the homogeneous background. It can be interpreted from these two profiles that both the GF and BF methods reduce noise but are not able to eliminate the noise completely and their performance are very much dependent on the acquisition durations. The performance of the NDF method is also very much dependent on the noise level. On the other hand, the GVOF method is able to remove noise for both of these acquisition durations and its performance is much less dependent on the noise level. The profiles are also flat in the homogeneous region for both the durations for the proposed method.

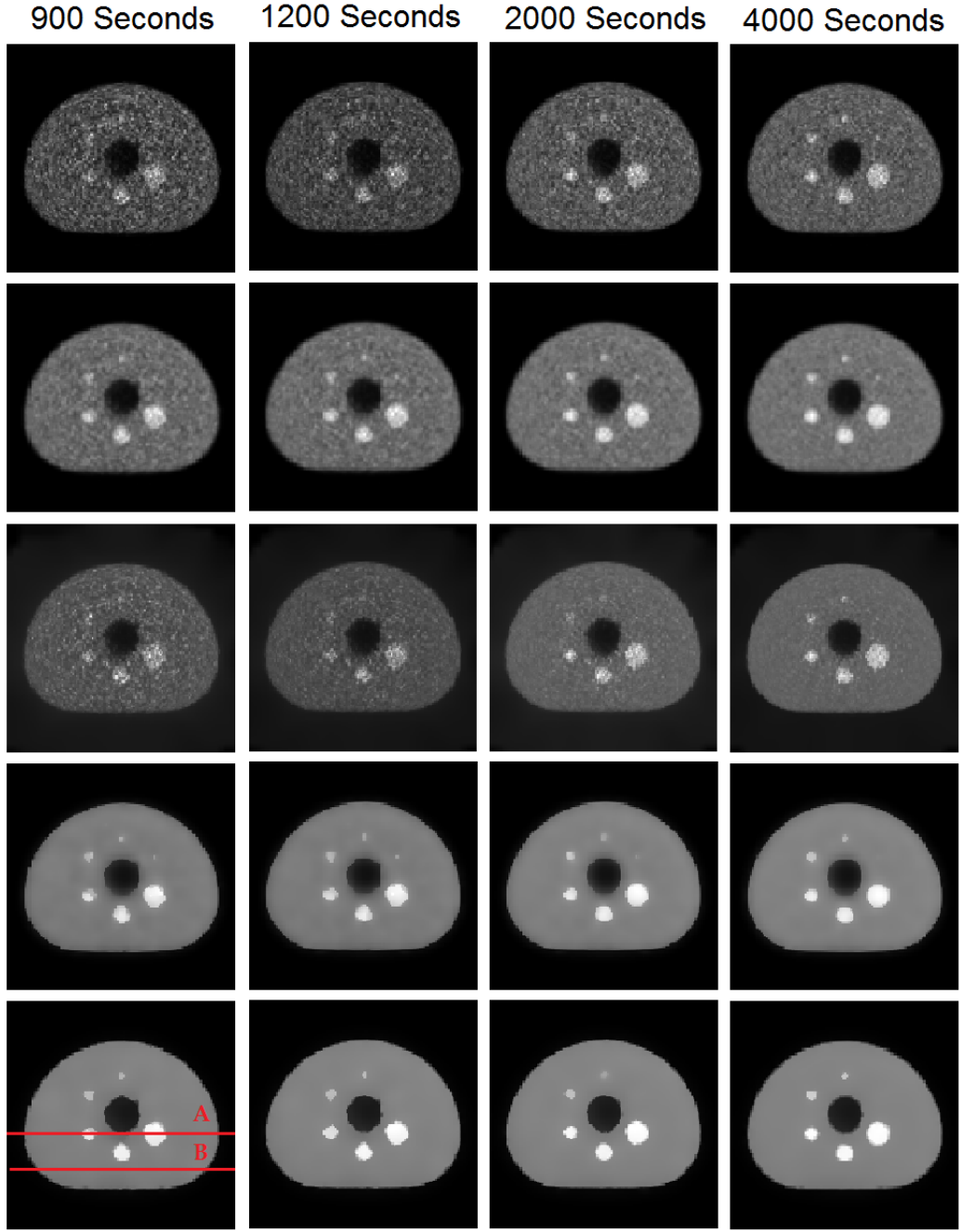

Figure 2: One PET slice of the NEMA phantom at contrast 2:1. Each column represents different scanning duration. First row: no filter; Second row: Gaussian Filter (GF) with 4 mm FWHM; Third row: Bilateral Filter (BF), Fourth row: Non-linear Diffusion Filter (NDF) and Fifth row: proposed method. Profiles through the horizontal red lines A and B are shown in Figure 3.

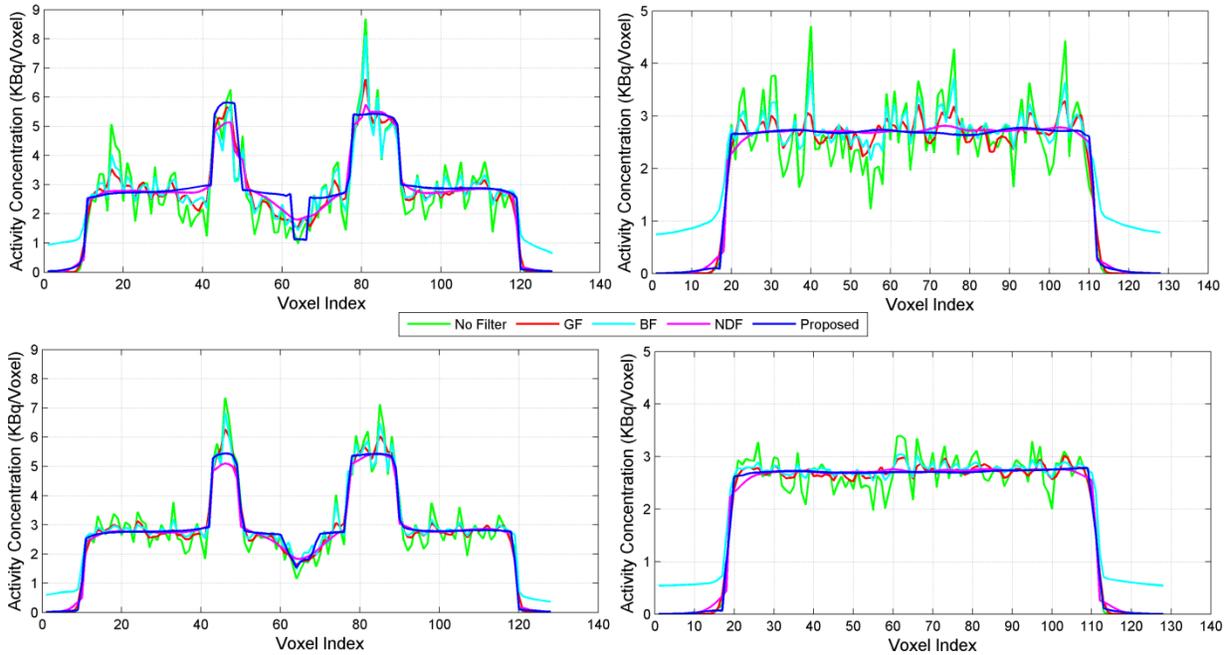

Figure 3: Horizontal profiles through the red line A (left column) and line B (right column) as shown in Figure 2 for all the four filtered images along with the original one. Top row: 900 seconds acquisition duration. Bottom row: 4000 seconds acquisition duration. For both the acquisition durations the performance of the proposed filter is the best.

The mean SNR and coefficient of variation (CoV), defined as the ratio of SD over mean of the SNR of five realizations for all methods and contrasts are shown in Table 2. The mean SNR increases and CoV decreases with the increase in acquisition durations. With the application of filtering, the mean SNR increases for all the methods. For GF, SNR is improved by 100% compared to no filter for 900 and 1200 seconds. However, for 2000 and 4000 seconds the improvement is less than 70%. GF also improves the CoV. At 2:1 contrast, no significant improvement is observed for the BF method compared to GF. However, at a higher contrast level at 4:1, 18% to 40% improvement is observed depending on the acquisition durations. Even at a higher contrast level, the BF method does not show any significant improvement in SNR over the GF method for 900 seconds acquisition duration. On the other hand, the CoV is approximately two times higher for BF compared to GF for all noise and contrast levels. Both the NDF and proposed method increase the SNR for all cases. The increase in the mean SNR for the NDF and proposed method is approximately 300% and 200% better than that of the no filter and GF methods respectively. For 900 and 2000 seconds durations, proposed method performs 10% better than NDF. However, CoV of SNR for proposed method is 100% lower compared to NDF for 900 and 2000 seconds acquisition durations indicating that the method is robust across acquisition duration. As the contrast and acquisition duration increases, differences in the CoV of the NDF method with the proposed method is reduced. Overall, the proposed method provides the highest mean SNR and the lowest CoV of SNR irrespective of the acquisition duration.

Table 2: Signal-to-Noise Ratio (SNR) for different acquisition durations and contrasts

|  |  | Contrast 2:1 | | | | Contrast 4:1 | | | |
| --- | --- | --- | --- | --- | --- | --- | --- | --- | --- |
|  |  | 900 Seconds | 1200 Seconds | 2000 Seconds | 4000 Seconds | 900 Seconds | 1200 Seconds | 2000 Seconds | 4000 Seconds |
| **No Filter** | Mean | 9.59 | 10.69 | 12.92 | 15.89 | 9.81 | 11.04 | 13.12 | 15.92 |
|  | CoV | 0.11 | 0.10 | 0.08 | 0.03 | 0.10 | 0.10 | 0.08 | 0.04 |
| **GF** | Mean | 18.73 | 19.88 | 21.99 | 24.86 | 18.76 | 19.92 | 21.86 | 24.61 |
|  | CoV | 0.070 | 0.069 | 0.042 | 0.008 | 0.07 | 0.07 | 0.05 | 0.03 |
| **BF** | Mean | 16.48 | 17.99 | 21.16 | 25.84 | 21.30 | 23.56 | 27.54 | 34.31 |
|  | CoV | 0.110 | 0.080 | 0.109 | 0.036 | 0.098 | 0.107 | 0.086 | 0.056 |
| **NDF** | Mean | 34.39 | 36.47 | 40.80 | 42.39 | 37.11 | 38.09 | 40.55 | 42.07 |
|  | CoV | 0.136 | 0.130 | 0.065 | 0.029 | 0.079 | 0.082 | 0.059 | 0.027 |
| **Proposed GVOF** | Mean | 37.51 | 39.02 | 40.59 | 42.43 | 39.10 | 40.50 | 42.73 | 43.99 |
|  | CoV | 0.067 | 0.065 | 0.039 | 0.012 | 0.036 | 0.041 | 0.047 | 0.033 |

CNR for all the methods for each sphere is shown in Figure 4. CNRs at contrast 4:1 are always higher compared to contrast 2:1 (on an average 275% higher) and it increases with the increase of scanning durations and size of the sphere. The improvement in CNR for the GF method ranges from 100% to 180% depending on the size of the sphere compared to no filter images with CNR values ranging from 0.79 to 4.84 with 2.26±1.03 (mean±SD) for contrast 2:1 and 0.91 to 12.50 with 5.04±2.88 for contrast 4:1 for no filter images. For GF images the CNR values range from 1.62 to 12.96 (5.77±2.92) and 2.55 to 34.61 (13.98±8.03) for contrast 2:1 and 4:1 respectively. The improvement for BF is less than 25% compared to GF for 2:1 contrast with values ranging from 0.86 to 13.82 (4.43±3.25). However, for contrast 4:1, BF improves CNR by an average of 198% with ranges from 5.48 to 123.69 (41.80±32.75). For 2:1 contrast, CNR is similar for both the NDF and proposed methods for 2000 and 4000 seconds acquisition durations and the improvement of CNR is 520% compared to the GF method (range 1.25 to 92.52 (46.40±29.65) for NDF and 0.79 to 96.80 (47.22±30.76) for GVOF). For 900 and 1200 seconds CNRs for the proposed method are 35% and 28% better than that of NDF (1.96 to 51.90 (25.56±15.33) for NDF and 0.86 to 66.24 (31.61±21.39) for GVOF). For contrast 4:1, proposed GVOF method provides 15% better CNR across different sizes and acquisition duration compared to NDF (range 53.43 to 290.66 (164.82±61.28) for NDF and 15.29 to 364.54 (190.21±93.96) for GVOF). Overall, images smoothed by the proposed filter provide the highest CNR irrespective of the acquisition duration and contrast.

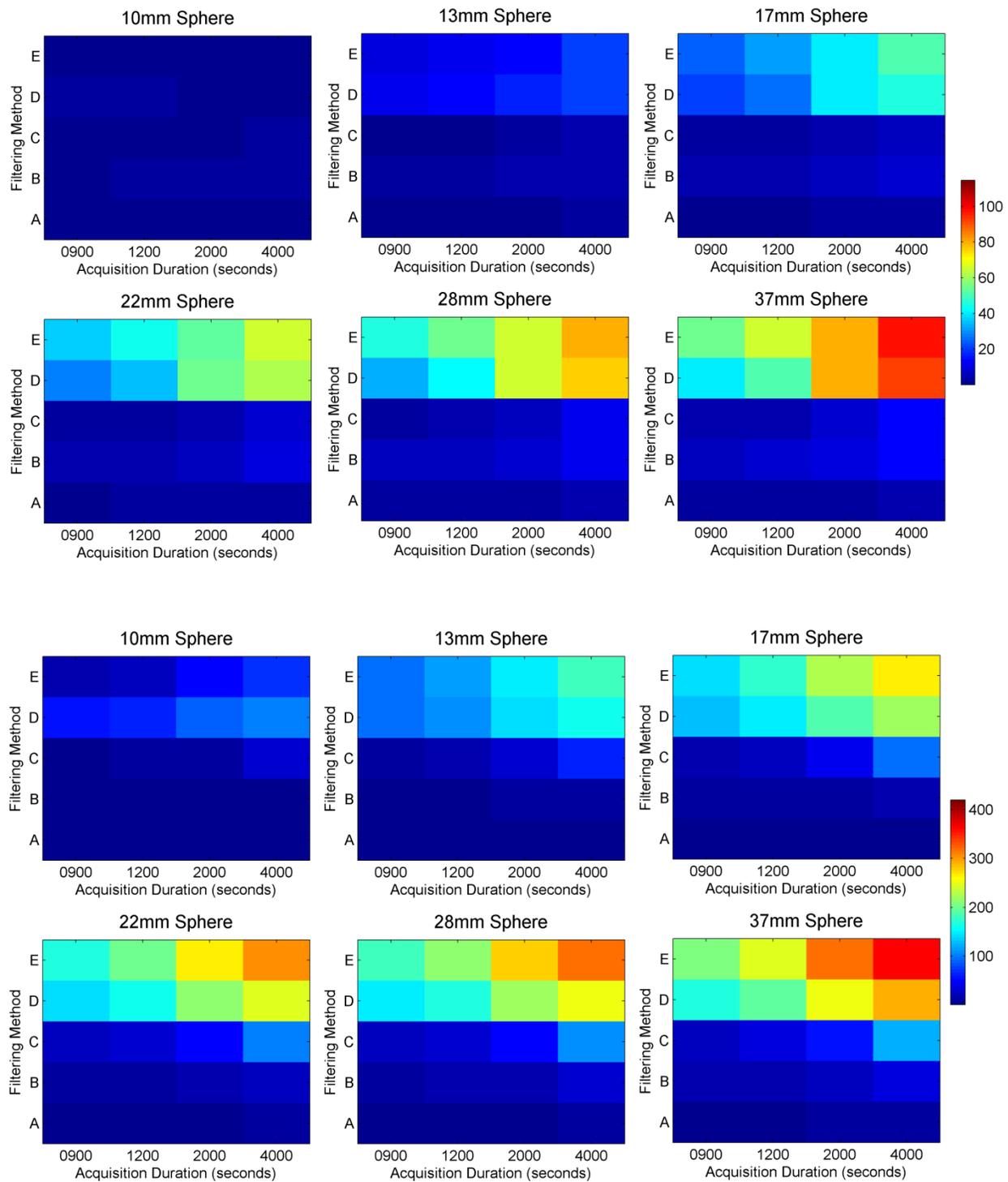

Figure 4: Contrast-to-Noise Ratio (CNR) for all the six spheres. The colour bar represents the value in each image. Top two rows for contrast 2:1 and bottom two rows for contrast 4:1. A - no filter, B – GF, C – BF, D – NDF and E – proposed method.

The mean and SD of the resolution of five realizations are shown in Table 3. Average resolution achieved by the OSEM reconstruction algorithm without any post reconstruction filtering is 6.40 mm. Application of GF degrades the mean resolution to 8.20 mm. However, SD of the resolution between different measurements is improved with GF. The mean resolutions over all acquisition durations provided by the BF and NDF methods are 6.45 mm and 6.75 mm respectively with the NDF method providing lower SD. The average resolution achieved by the proposed method is 5.8 mm and is similar across all noise levels. The standard deviation is also the lowest compared to all the other methods.

Table 3: Resolution (FWHM) in mm

|  |  | Contrast 2:1 | | | | Contrast 4:1 | | | |
| --- | --- | --- | --- | --- | --- | --- | --- | --- | --- |
|  |  | 900 Seconds | 1200 Seconds | 2000 Seconds | 4000 Seconds | 900 Seconds | 1200 Seconds | 2000 Seconds | 4000 Seconds |
| **No Filter** | Mean | 6.57 | 5.16 | 6.59 | 6.52 | 4.97 | 6.74 | 6.99 | 7.29 |
|  | SD | 2.46 | 2.00 | 1.66 | 2.33 | 1.26 | 1.63 | 0.85 | 1.39 |
| **GF** | Mean | 7.13 | 7.85 | 7.88 | 9.26 | 8.63 | 7.80 | 7.82 | 8.86 |
|  | SD | 0.37 | 0.47 | 1.00 | 0.77 | 1.96 | 1.33 | 0.52 | 0.91 |
| **BF** | Mean | 5.96 | 5.92 | 7.24 | 6.89 | 5.86 | 6.64 | 6.56 | 6.67 |
|  | SD | 1.51 | 1.95 | 2.23 | 0.75 | 2.68 | 1.94 | 0.65 | 0.81 |
| **NDF** | Mean | 6.65 | 6.55 | 6.19 | 5.77 | 7.45 | 8.29 | 7.42 | 7.25 |
|  | SD | 1.02 | 1.36 | 0.84 | 0.48 | 1.27 | 1.64 | 0.60 | 0.16 |
| **Proposed GVOF** | Mean | 5.47 | 5.66 | 5.30 | 5.88 | 5.86 | 5.84 | 6.01 | 6.21 |
|  | SD | 0.37 | 0.51 | 0.64 | 0.27 | 0.34 | 0.16 | 0.21 | 0.21 |

The maximum activity (analogous to $SUV_{max}$) of all the five realizations for contrast 2:1 and 4:1 for all spheres and filtering methods are shown in Figure 5. The mean and standard deviation (as a measure of variance) for five realizations along with the measured activity that was used to fill the spheres are also shown in the figure. Figure 6 shows the percentage bias in comparison to the actual activity. Spheres with diameter less than 2 cm (i.e., 2-3 times smaller the than resolution of the scanner) are subject to higher level of partial volume effects and accurate correction for PVE is vital in such cases [35]. Without accurate PVE correction they may provide biased result.

Because of this reason, to compare the performance of different filtering methods bias for the spheres less than 2 cm diameter is analyzed separately.

The mean of the five realization of the maximum values of No Filter an BF methods for 10 and 13 mm spheres range from 7.64 to 13.74 and 6.4 to 12.47 across different acquisition durations with corresponding bias of -1.75 to 70.32% (20.31±22.06%) and -0.42 to -55.17% (2.47±21.08) respectively. Without PVE correction, the estimated value should ideally be lower than that of the measured or true value as in the case of GF (-20.31±10.43%), NDF (-30.28±12.28) and GVOF (-40.47±12.32). However, since the BF method is not able to remove noise for completely, the estimated maximum values match with the measured value. In such case, it not the method that is accurate rather it is the noise that randomly effects the estimation of the maximum value.

For 17mm sphere, the performance of the GF method is the best in terms of bias (3.71±4.10%) followed by NDF (-13.75±5.95%), GVOF (-17.20±1.92%), BF (37.10±19.06%) and no filter (51.71±20.47%) respectively for contrast 2:1. For contrast 4:1, the differences between bias for GF, NDF and GVOF reduce significantly. Mean±SD of the bias values for these methods are 6.67±2.46, 1.94±3.50 and 10.55±0.41 respectively with the GVOF being the least sensitive to the acquisition duration. No filter (55.52±20.61) provides the worst bias followed by the BF (44.19±22.13).

For the biggest three spheres at both contrast levels, the average of the estimated maximum activity and hence the percentage bias are very much dependent on the noise levels for all methods (percentage bias range: No Filter (19.29 to 124.69%), GF (6.78 to 33.82%), BF (13.22 to 111.52%) and NDF (-0.56 to 28.68%)) except the proposed method (0.0 to -7.06%). The mean±SD of the percentage bias for these three spheres are 85.87±30.37%, 21.02±7.45, 72.27±30.94, 9.30±12.85 and -0.28±3.41% for No Filter, GF, BF, NDF and GVOF methods respectively for contrast 2:1. For contrast 4:1 the values are 65.37±21.41, 15.79±6.33, 50.18±22.10, 11.19±8.20 and -2.51±2.57.

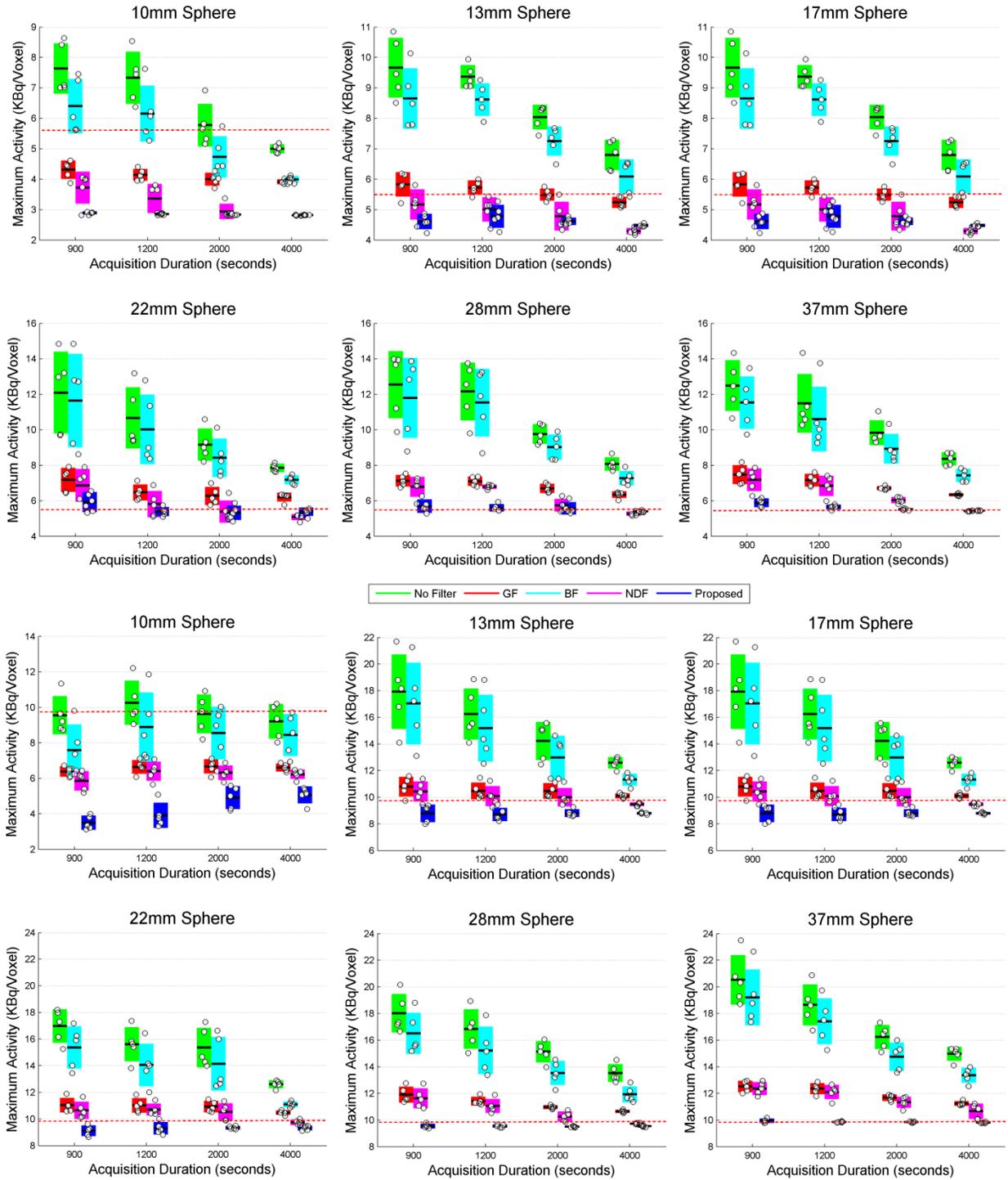

Figure 5: Box plot with mean and standard deviation for maximum activity in each sphere for contrast 2:1 (top two rows) and contrast 4:1 (bottom two rows). Each circle represents maximum activity in each sphere. The dashed red line is the activity that has been used to fill the spheres.

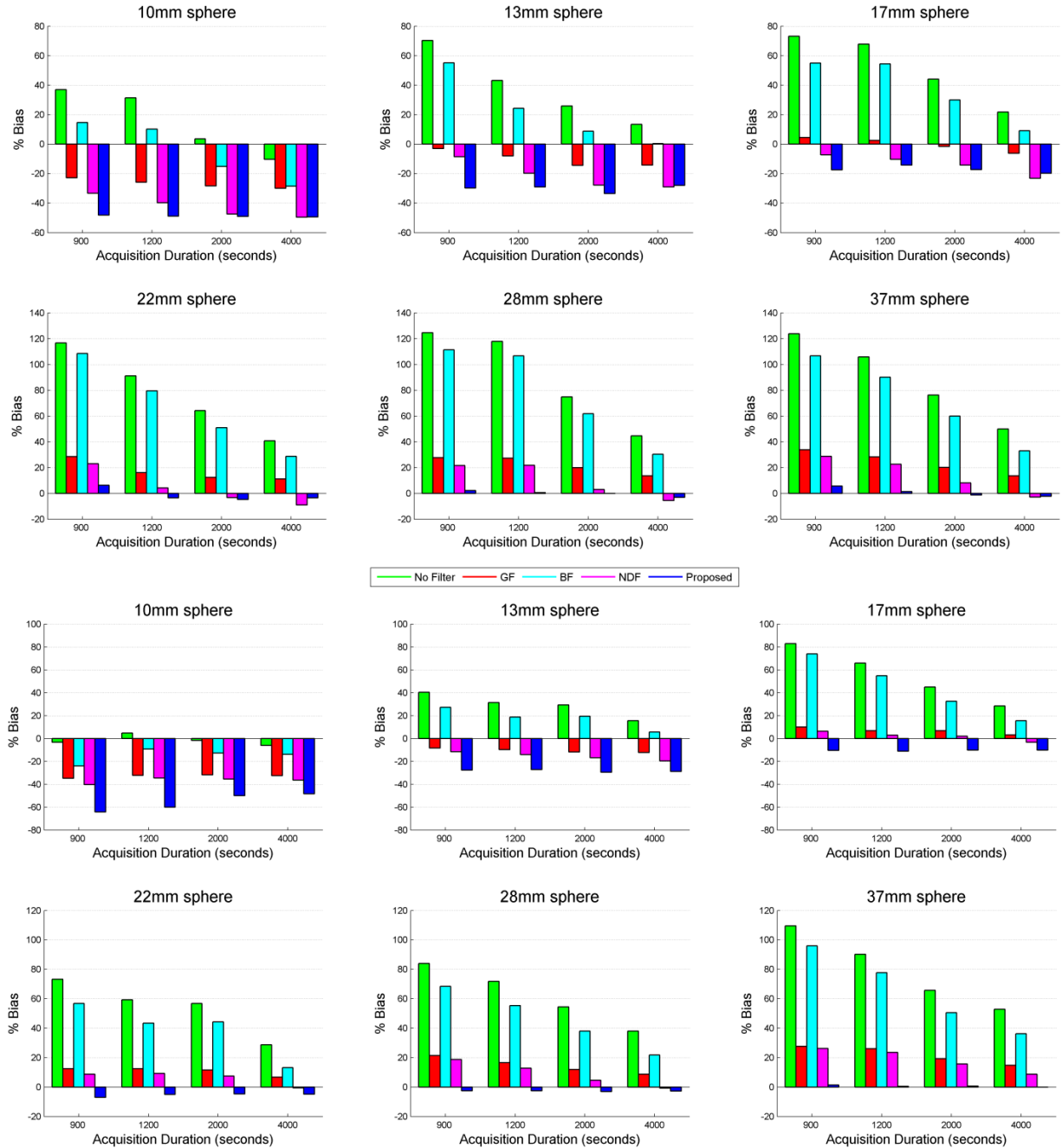

Figure 6: Percentage bias for contrast 2:1 (top two rows) and contrast 4:1 (bottom two rows)

Reproducibility between the highest and lowest maximum activity for each sphere for each method is shown in Figure 7. With the application of filtering, reproducibility increases for all methods except BF. Reproducibility for all methods are dependent on the contrast, acquisition durations and size.

For the smallest two spheres for both the contrast levels, the highest reproducibility provided by the GF with mean±SD of 11.82±4.81 followed by the NDF (18.11±9.04), GVOF (18.79±13.90) no filter (27.13±9.12) and BF (36.67±11.58) method with difference of 35%, 37%, 57% and 68% respectively.

For the biggest four spheres (≥ 17mm), the percentage reproducibility for no filter images varies from 6.79 to 42.09% for contrast 2:1 (21.78±11.12) and 5.84 to 42.40% (19.01±8.67) for contrast 4:1 based on the acquisition duration. For the GF, the values are 2.62 to 20.45% (11.07±5.23) for contrast 2:1 and 2.63 to 17.51% (8.39±4.37) for contrast 4:1. Reproducibility of the BF method is the worst with values vary from 8.14 to 53.02% (27.09±13.28) for contrast 2:1 and 4.98 to 40.09% (23.34±10.86) for contrast 4:1. The NDF method improves reproducibility compared to no filter and BF but degrades compared to GF for with values ranging from 0.84 to 32.23% (16.91±9.94) for contrast 2:1 and 1.34 to 19.49 (11.19±5.06) for contrast 4:1. The GVOF method provides the best reproducibility with values 1.11 to 20.66% (10.54±6.63) and 1.22 to 16.11% (5.48±4.90) for contrast 2:1 and 4:1 respectively. The reproducibility improvements by the GVOF method are 106%, 5%, 157% and 60% for contrast 2:1 and 247%, 53%, 326% and 104% for contrast 4:1 compared to no filter, GF, BF and NDF method respectively.

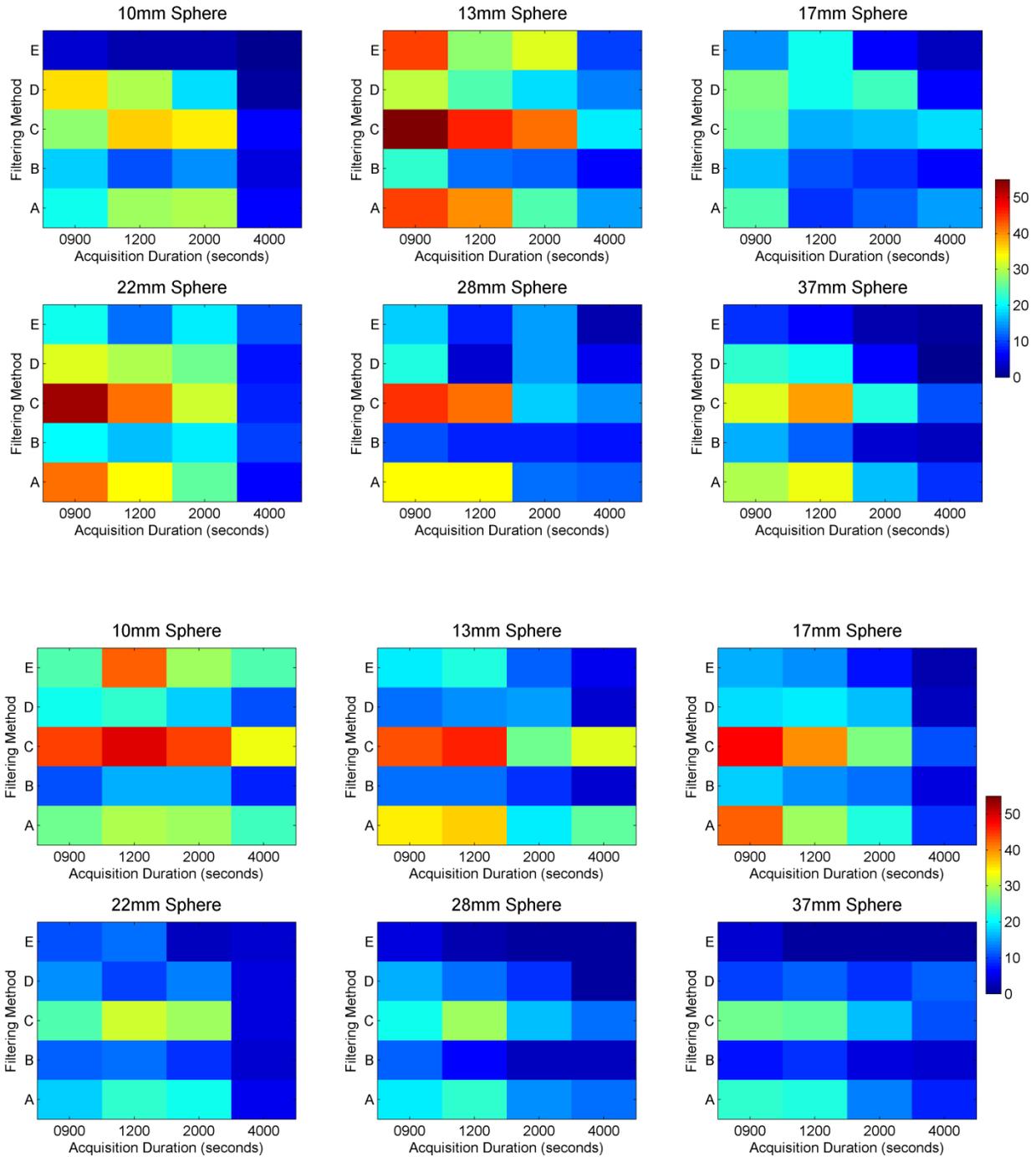

Figure 7: Reproducibility for contrast 2:1 (top two rows) and contrast 4:1 (bottom two rows). A - no filter, B – GF, C – BF, D – NDF and E – proposed method.

**DISCUSSION**

The proposed GVOF filtering method takes into account the orientation of the gradient direction within a specified window along with the gradient magnitude to calculate diffusion coefficient. This approach provides two advantages over other approaches – 1) detection of features with both low and high contrast and 2) removal of features due to noise. The first advantage removes the requirement of recently proposed filtering method that requires combination of two separate steps for low and high contrast [23]. The GVOF method is data driven and does not require prior training. The performance of the method was compared with other most used and state-of-the-arts method using several quantitative measures, e.g., SNR, CNR, resolution, percentage bias of the maximum intensity and reproducibility along with the qualitative visual assessment. Image noise and lesion size and contrast were also varied to assess the robustness of the new filter against statistical fluctuations. The robustness is vital if the filter is going to be used with images acquired for response assessment where the aforementioned parameters are subjected to change due to treatment.

The SNR and CNR are the highest for the proposed method across different contrasts and sizes. The method is the least sensitive to the change in acquisition duration which represent the level of noise in the image. Only for the 10 mm sphere, the CNR of NDF is higher to that of GVOF. This is due to the fact that GVOF works by taking the average of the gradient direction within a specified window size which results in the reduction of contrast for very small objects. The average CNR is multiple times higher than previously reported even at a low contrast of 2:1 (39.42 vs 19.46) [23].The CNR improvement ratio compared to the images with no filter is also several times higher than recently proposed deep learning based denoising approach [24].

The edge preserving filtering methods BF and NDF preserve the resolution of the images at all noise levels with the performance of the former being very much dependent on the noise level. Both the method also reduce the variability compared to the images with no filter. The proposed filtering method not only provides the best mean resolution with the lowest variability compared to all other methods and performs robustly across the different noise levels but also improves the resolution by approximately 20% compared to the original image.

Lesion in PET are subjected to higher PVE if the diameter of the lesion <3 times of the resolution of the image. Because of this reason for quantitative estimation of the maximum activity partial volume correction needs to be deployed for lesion with diameter <2cm [36]. Without PVE correction, the estimated activity ideally should be less than that of the actual measurement as in the case of GVOF for the smallest three spheres (diameter < 2 cm). For the biggest three spheres (diameter > 2 cm) which is typically the criteria for inclusion of patient for clinical evaluation [37], the GVOF method estimates the most accurate radioactivity concentration across different noise levels and contrasts and much lower than the recently proposed method (average percentage bias of 1.39% vs 27.5%) [23].

The improvement of bias between two contrasts for the GVOF method indicates that for low contrast the spill out effect is higher when the background activity is lower than the lesion activity. Though the performance of the proposed GVOF method is not far apart from the GF and NDF for 17mm sphere, with accurate PVE the performance can further be improved. It is reasonable to anticipate that with accurate PVE correction, the proposed method will be able to provide accurate estimation of the radioactivity concentration even for lesion having diameter <

2cm. However, it has been reported that for a very small lesion (< 10mm diameter) the PVE correction method may not be accurate [38]. The reproducibility of the proposed method is also the highest compared to all the other methods.

The robustness of the GVOF method against noise also makes it a suitable filter to be used in dynamic PET where the early frame durations are generally shorter and hence noisy to capture the distinctive features of the time activity curve, e. g, peak in the blood time activity curve [39]. Since, the proposed method is the least sensitive to the changes in size, contrast and noise, it can be useful for response assessment as well as for tracer with differential uptake.

**CONCLUSION**

A novel non-linear diffusion filtering method to remove noise from PET images is presented in this paper overcoming limitations of other methods. The method utilizes combined information provided by both gradient magnitude and gradient vector direction. The proposed GVOF method not only performs robustly across changes in image contrasts and noise but also increase the image resolution. Along with the higher SNR, CNR and resolution, the proposed method provides the best quantitative accurate results (percentage bias, variance and reproducibility) for lesions that do not require any partial volume correction, i.e., lesion with size greater than three times of the PET intrinsic resolution. Moreover, the method does not require to adjust parameters (e.g., iteration number, $\kappa$ values etc.) for different statistical settings arising from changes in size, contrast and noise due to response to therapy or differential uptake of radiotracers. This makes the proposed method an strong alternative to Gaussian filter that are currently used in the state-of-the-art clinical and preclinical PET systems to provide smoothed images for the purpose of both diagnosis and response assessment. On the other hand, because of its capability to provide accurate quantitative measurements irrespective of the SNR, it can also be effective in reduction of radioactivity dose which will require further investigation.

## Data availability

The data related to the findings of this study are available from the authors on reasonable request.

## Code availability

All software code including a single organized script written using MATLAB 2014 is freely shared under the GNU General Public License on the GitHub website at: https://github.com/mtamal/GVOF. The access to the code is protected as a part of patent restriction requirement.

## Applicable Funding Source

The author extend his appreciation to the Deputyship for Research & Innovation, Ministry of Saudi Arabia for funding this research work.